\documentclass[twocolumn,showpacs,preprintnumbers,amsmath,amssymb]{revtex4}
\usepackage{graphicx}
\usepackage{dcolumn}
\usepackage{bm}

\begin{document}

\title{Exact results for nonlinear ac-transport through a resonant level model}

\author{P. Wang}
\email{pei.wang@physik.lmu.de}
\author{M. Heyl}
\author{S. Kehrein}
\affiliation{
Physics Department, Arnold Sommerfeld Center for Theoretical Physics, and Center for NanoScience, \\
Ludwig-Maximilians-Universit\"at, Theresienstrasse 37, 80333 Munich, Germany
}

\date{\today}

\begin{abstract}
We obtain exact results for the transport through a resonant level model 
(noninteracting Anderson impurity model) for rectangular voltage bias as a function of time. 
We study both the transient behavior after switching on the tunneling at time $t=0$
and the ensuing steady state behavior. Explicit expressions are obtained for the
ac-current in the linear response regime
and beyond for large voltage bias. Among other effects, we observe current ringing
and PAT (photon assisted tunneling) oscillations. 
\end{abstract}

\maketitle

\section{Introduction}

The recent advances in nanotechnology created a lot of interest in transport through
correlated quantum impurities. While the linear response regime essentially probes
the ground state properties of the system, transport beyond the linear response 
regime explores genuine non-equilibrium quantum many-body phenomena. 
However, theoretical calculations beyond the linear response regime are 
challenging since the steady state cannot be constructed via a variational principle
like equilibrium states. Even for dc-bias
only recently exact numerical methods have been developed that
permit such investigations for interacting systems, notably the time-dependent numerical 
renormalization group \cite{AndersPRL101}, Monte Carlo methods \cite{Schmidt2008,Millis2009},
and the time-dependent density matrix renormalization group \cite{WhitePRL2004,Schmitteckert2004}. 
Some of the analytical methods that have been applied successfully are perturbative Keldysh
calculations \cite{Fuji2003}, extensions of the renormalization group \cite{Rosch2001,Schoeller2009}, 
flow equations \cite{Kehrein_Kondo}, 
and generalizations of NCA (non-crossing approximation) to non-equilibrium \cite{nordlander,Meir93}. 
A comparative review of theoretical methods can be found in Ref.~\cite{FabianNewJPhys} 

For ac-bias beyond the linear response regime still
much less is known since, e.g., the numerical methods cannot easily be generalized
to time-dependent bias. Interesting ac-phenomena are for example
the photon assisted tunneling effect (PAT) \cite{tien} that has been observed in experiments~\cite{kouwenhoven94},
or the "current ringing" after a step-like bias puls \cite{wingreen94}.
Non-equilibrium Green's function methods can be employed~\cite{wingreen93,wingreen94,maciejko06}
when the correlation effects are not too strong.
In the strongly correlated regime of the Kondo model the non-crossing approximation was found to be reliable 
\cite{nordlander,plihal,Goker08,nordlander2000}. At a specific point of the two-lead Kondo model it can be solved 
exactly~\cite{hershfield96}, which permits exact results for
the current in the steady state~\cite{hershfield95}, after a rectangular pulse~\cite{hershfield00} or under sinusoidal bias~\cite{hershfield96}. Unfortunately, this special point is not generic for a Kondo impurity that can
be derived from an underlying Anderson impurity model, which is experimentally the most relevant situation.

In this paper we study the response of a resonant level model (noninteracting Anderson impurity model) under rectangular 
ac voltage bias after switching on the tunneling at time $t=0$. 
We derive exact analytical results for the transient and steady-state current by diagonalizing the Hamiltonian. 
This exact solution 
contains both dc- and ac-bias in and beyond the linear response regime. 
While dc-results and ac-results with sinusoidal bias have been obtained previously in the literature \cite{wingreen94},
rectangular ac-driving beyond the linear regime seems not to have been studied before.
Besides being experimentally relevant, our results are also helpful
for exploring the various crossovers in this important model and serve as an exact benchmark for future work.

\section{Model and diagonalization}

The resonant level model coupled to two leads is defined by the following Hamiltonian
\begin{eqnarray} \nonumber
H&=& \sum_{k\alpha} \epsilon_k c^\dag_{k\alpha} c_{k\alpha} + \sum_{k\alpha} \frac{g}{\sqrt{2}} (c^\dag_{k \alpha} d+ h.c.) ,
\end{eqnarray}
where $\alpha=L,R$ denotes the leads. The spin index can be omitted since the model is non-interacting and we work with spinless fermions. All energies
are measured with respect to the single-particle energy of the impurity orbital ($\epsilon_d \equiv 0$).
We take a wide band limit with a linear dispersion relation, $\epsilon_k = k\eta $, where $\eta$ denotes the level spacing and
$k$ an integer number. The hybridization is defined by $\Gamma= \rho\pi g^2$ where $\rho=1/\eta$. The impurity orbital
spectral function in equilibrium is then given by
\begin{equation}
\rho_d(\epsilon)=\frac{\Gamma}{\pi(\epsilon^2+\Gamma^2)} 
\end{equation}

Our strategy to obtain exact results is to first
diagonalize the discretized Hamiltonian and to then take the thermodynamic limit $\eta\rightarrow 0$.
We introduce the hybridized basis $c_{s} = \sum_k \frac{g}{\epsilon_s - \epsilon_k} B_s c_{k + } + B_s d$.
It is then straightforward to diagonalize the Hamiltonian
\begin{eqnarray}
 H =\sum_k \epsilon_k c^\dag_{k - } c_{k -} + \sum_s \epsilon_s c^\dag_{s} c_{s},
\end{eqnarray}
where $c_{k \pm }= \frac{1}{\sqrt{2}} (c_{k L} \pm c_{kR})$. The inverse transformation is
$d= \sum_s B_s c_{s }$ and
$c_{k+}=\sum_s \frac{g}{\epsilon_s - \epsilon_k} B_s c_{s}$.  The eigenvalues are determined 
as solutions of the equation
\begin{equation} 
\frac{\epsilon_s}{g^2} = \frac{\pi}{\eta} \cot \frac{\pi \epsilon_s}{\eta}.
\end{equation}
In the thermodynamic limit
\begin{equation}
B^2_s = \frac{g^2}{\epsilon_s^2+\Gamma^2}.
\end{equation}
From the diagonalization one also derives the following set of equations
 \begin{eqnarray}\nonumber \label{complete}
 \sum_s B^2_s &=&1, \\ \nonumber
 \sum_s \frac{g^2 B^2_s}{(\epsilon_s- \epsilon_k)^2} &=&1, \\ \nonumber
 \sum_s \frac{B^2_s}{\epsilon_s -\epsilon_k} &=&0, \\ \nonumber
 \sum_s \frac{B_s^2}{(\epsilon_s - \epsilon_k)(\epsilon_{s}-\epsilon_{k'})} &=& 0 , k'\neq k.
 \end{eqnarray}
which will be important for calculating various summations below.
 
An ac voltage bias leads to time-dependent potentials $u_a(t)$ in the leads
and the Hamiltonian takes the form
\begin{equation}
H= \sum_{k\alpha} (\epsilon_k-u_\alpha(t)) c^\dag_{k\alpha } c_{k\alpha} + \sum_{k\alpha} \frac{g}{\sqrt{2}} (c^\dag_{k \alpha} d
+  h.c.).
\end{equation}
We suppose that initially (at time $t<0$) the left and right lead 
chemical potential are the same, $\mu_L = \mu_R =\mu$, the hybridization is switched off
and that there is no electron in the dot, $n_d=0$. 
At time $t=0$ the hybridization is switched on and  
a rectangular voltage bias with period~$2T$ (see Fig.~2) is applied: 
$u_R(t) = -u_L(t)= V /2$ for $2N T<t<(2N+1)T$ and $u_R(t) = -u_L(t)=- V /2$ for 
$(2N+1) T<t<2(N+1) T$.\cite{levelV} $\mu$ therefore gives the energy difference of the
impurity orbital to the "average" Fermi energy of the leads for time~$t>0$ (Fig.~2).

The current operator is defined as
\begin{eqnarray}
I_{\alpha} =s_\alpha e \frac{dN_{\alpha }}{dt} =\frac{iges_\alpha}{\sqrt{2}} \sum_k(d^\dag c_{k\alpha} - c^\dag_{k \alpha} d ),
\end{eqnarray}
where $N_\alpha$ denotes the total number of electrons in lead~$\alpha$ and
$s_L\stackrel{\rm def}{=}1$, $s_R\stackrel{\rm def}{=}-1$.

\begin{figure}
\begin{center}
\includegraphics[width=0.45\textwidth]{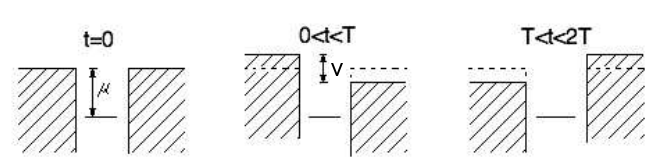}
\caption{A schematic diagram of our model: A step-like voltage bias is applied to the two leads coupled to the quantum dot.}
\end{center}
\label{fig_1}
\end{figure}

In the first half period $2N T<t<(2N+1)T$ the Hamiltonian is
\begin{eqnarray}\nonumber
H_a &=& \sum_{k} (\epsilon_{k}+ \frac{V}{2}) c^\dag_{k L} c_{kL}
+ \sum_{k} (\epsilon_{k}- \frac{V}{2}) c^\dag_{k R } c_{kR } \\
&& + \sum_{k\alpha} \frac{g}{\sqrt{2}} (c^\dag_{k \alpha } d + h.c.). 
\end{eqnarray}
Because the dispersion relation
is linear and $k$ runs from $-\infty$ to $\infty$ (wide band limit),
we can simply relabel the fermion operators, $c_{k\alpha } = \tilde c_{k+\frac{s_\alpha V}{2\eta}, \alpha }$. 
The potentials in the leads are eliminated 
by this transformation and the Hamiltonian can be diagonalized as before:
$ H_a=\sum_s \epsilon_s  a^\dag_{s} a_{s} + \sum_k \epsilon_k a^\dag_{k-}
a_{k-}$, where  $a_{s} =\sum_k \frac{g B_s}{\epsilon_s - \epsilon_k}a_{k+ } +B_s d $ and
$a_{k \pm }= \frac{1}{\sqrt{2}} (c_{k-\frac{\rho V}{2},L} \pm c_{k+\frac{\rho V}{2},R})$. 
Similarly, in the second half period $(2N+1) T<t<2(N+1) T$ the Hamiltonian is diagonalized as
$H_b = \sum_s \epsilon_s  b^\dag_{s} b_{s} + \sum_k \epsilon_k b^\dag_{k-}
b_{k-}$, where $b_{s} = \sum_k \frac{g B_s}{\epsilon_s - \epsilon_k}b_{k+} +B_s d  $ and $b_{k\pm }= \frac{1}{\sqrt{2}} (c_{k+\frac{\rho V}{2},L} 
\pm c_{k-\frac{\rho V}{2},R})$.

In the Heisenberg picture the current operator at time $t=2NT+\tau, \tau\in [0,T]$ (first half period) can be expressed as
\begin{eqnarray}
I_\alpha (t)= (e^{iH_aT} e^{iH_b T})^N e^{iH_a\tau} I_\alpha  e^{-iH_a \tau} (e^{-iH_bT}e^{-iH_aT} )^N,
\end{eqnarray}
and in the second half period ($t=(2N+1)T+\tau, \tau\in [0,T]$) 
\begin{eqnarray}\nonumber
I_\alpha (t)&=& (e^{iH_aT} e^{iH_b T})^{N} e^{iH_a T}e^{iH_b\tau} I_\alpha  \\ && \times e^{-iH_b \tau} e^{-iH_a T}(e^{-iH_bT}e^{-iH_aT} )^{N}.
\end{eqnarray}
To find $I_\alpha(t)$ we first calculate the time evolution of the single fermion operator $d^\dag$ and $c^\dag_{k\alpha}$ under $H_a$ or $H_b$
by expressing $d^\dag$ and $c^\dag_{k\alpha}$ in the hybridized basis, next
applying the diagonal time evolution and finally transforming back to the original basis. 
The calculation is straightforward but one needs to pay attention when encountering
summations with respect to the eigenenergies $\epsilon_s$.
In the thermodynamic limit the summation can be transformed into an integral when there is no pole in the integrand, e.g., 
$\sum_s B^2_s e^{-i \epsilon_s t} =\int d\epsilon_s \rho B^2_s e^{-i\epsilon_s t}=  e^{-\Gamma t} $. If there are poles in the integrand we first calculate the time derivative to get rid of the pole terms. 
Key formulas are
\begin{eqnarray}
\sum_s \frac{B^2_s e^{-i\epsilon_s t}}
{\epsilon_s - \epsilon_k} &=& \frac{e^{-i \epsilon_k t}- e^{-\Gamma t}}{\epsilon_k + i\Gamma} \ ,\\
\nonumber
\sum_s \frac{B^2_s e^{-i\epsilon_s t}}
{(\epsilon_s - \epsilon_k)^2} &=& (\frac{1}{g^2} + \frac{-1-( i\epsilon_k -\Gamma)t}{(\epsilon_k+i\Gamma)^2}) e^{-i \epsilon_k t} \\
&& + \frac{e^{-\Gamma t}}{(\epsilon_k+i\Gamma)^2} \ .
\end{eqnarray}
By using these two formulas we get
\begin{eqnarray}
&& e^{iH_{(a,b)}T} d^\dag e^{-iH_{(a,b)}T} = e^{- \Gamma T} d^\dag+\sum_{k\alpha} \frac{g}{\sqrt{2}} W^{(a,b) }_{k\alpha}
c_{k\alpha}^\dag \\ && \nonumber
e^{iH_{(a,b)}T} c^\dag_{k\alpha} e^{-iH_{(a,b)}T} = \frac{g}{\sqrt{2}} W^{(a,b) }_{k\alpha} d^\dag \\ && \nonumber
+ \sum_{k'\alpha'} (\frac{g^2(W^{(a,b) }_{k\alpha}  -W^{(a,b) }_{k'\alpha'} )}{2(\epsilon_{k\alpha}^{(a,b)} - 
\epsilon_{k'\alpha'}^{(a,b)})}  + \delta_{\alpha,\alpha'}\delta_{k,k'} e^{i \epsilon_{k\alpha}^{(a,b)}T } ) c_{k'\alpha'}^\dag, \\
\end{eqnarray}
where $W^{(a,b) }_{k\alpha}(T) =  \frac{e^{i\epsilon^{(a,b)}_{k\alpha}T} -e^{-\Gamma T}}{\epsilon_{k\alpha}^{(a,b)}- i\Gamma}$, $\epsilon^a_{k L}=\epsilon^b_{k R} =\epsilon_k + V/2$ and $\epsilon^a_{k R}=\epsilon^b_{k L} = \epsilon_k - V/2$.
Employing this formula twice gives the evolution over a full period:
\begin{widetext} 
\begin{eqnarray}\nonumber
 e^{iH_aT}e^{iH_bT} d^\dag e^{-iH_bT} e^{-iH_aT} &=& e^{-2\Gamma T} d^\dag +\sum_{k\alpha} \frac{g}{\sqrt{2}} D^{a}_{k\alpha} c_{k\alpha}^\dag,   \\
  e^{iH_aT}e^{iH_bT} c^\dag_{k\alpha} e^{-iH_bT} e^{-iH_aT} &=& \frac{g}{\sqrt{2}}
 D^{b}_{k\alpha} d^\dag +\sum_{k'\alpha'} (K_{k'\alpha',k\alpha}+ \delta_{k,k'} \delta_{\alpha, \alpha'} e^{2i \epsilon_k T}) c_{k'\alpha'}^\dag  ,
\end{eqnarray}
where 
\begin{eqnarray}\nonumber
D^{(a,b) }_{k\alpha} &=& e^{i \epsilon^{(a,b)}_{k\alpha}T }W^{(b,a)}_{k\alpha}+ e^{-\Gamma T} W^{(a,b)}_{k\alpha}, \\ K_{k'\alpha',k\alpha} &=& 
e^{i\epsilon^a_{k'\alpha'}T} \frac{g^2 (W^b_{k'\alpha'} - W^b_{k\alpha})}{2(\epsilon^b_{k'\alpha'} - \epsilon^b_{k\alpha})} +e^{i\epsilon^b_{k\alpha}T} 
\frac{g^2 (W^a_{k'\alpha'} - W^a_{k\alpha}) }{2(\epsilon^a_{k'\alpha'} - \epsilon^a_{k\alpha})}  + \frac{g^2}{2} W^{a}_{k'\alpha'} W^{b}_{k\alpha} .
\end{eqnarray}
 \end{widetext}
We perform the summation over $k$ by transforming it into an integral and then employing the residue theorem. 
Applying the above formula recursively $N$ times yields
\begin{widetext}
\begin{eqnarray}\nonumber
 (e^{iH_aT}e^{iH_bT})^N d^\dag (e^{-iH_bT} e^{-iH_aT})^N &=& e^{-2 N \Gamma T} d^\dag + \sum_{k\alpha}\frac{g}{\sqrt{2}} 
D^{a}_{k\alpha} \gamma_N(k) c_{k\alpha}^\dag,   \\ \nonumber
  (e^{iH_aT}e^{iH_bT})^N c^\dag_{k\alpha} (e^{-iH_bT} e^{-iH_aT})^N &=&\sum_{k\alpha} \frac{g}{\sqrt{2}}
 D^{b}_{k\alpha} \gamma_N(k) d^\dag +\sum_{k'\alpha'} (\alpha_N(k',k) K_{k'\alpha',k\alpha} \\ && + \delta_{k,k'} \delta_{\alpha, \alpha'} 
e^{2Ni \epsilon_k T}+ \frac{g^2}{2}
 \beta_N(k',k) D^a_{k'\alpha'} D^b_{k\alpha} ) c_{k'\alpha'}^\dag ,
\end{eqnarray}
\end{widetext}
where $\alpha_0=\beta_0=\gamma_0=0$ and the recursion relations are
\begin{eqnarray}\nonumber
\alpha_{N+1}(k',k) &=& \alpha_N(k',k) e^{2i\epsilon_{k'}T}+e^{2Ni\epsilon_k T}, \\ \nonumber
\beta_{N+1}(k',k) &=& \beta_N(k',k) e^{2i\epsilon_{k'}T} + \gamma_N(k), \\ \nonumber
\gamma_{N+1}(k) &=& \gamma_N(k) e^{-2\Gamma T} + e^{2Ni \epsilon_kT}.
\end{eqnarray}
It is easy to find 
\begin{eqnarray}
\alpha_N&=& \frac{e^{2Ni\epsilon_k T}-e^{2Ni\epsilon_{k'}T}}{e^{2i\epsilon_k T}- e^{2i\epsilon_{k'}T}} \\
\gamma_N &=& \frac{e^{2Ni\epsilon_k T}-e^{-2N\Gamma T}}{e^{2i\epsilon_k T}- e^{-2\Gamma T}}
\end{eqnarray}
In the first half period the current evaluates to
\begin{eqnarray}\label{current_a}\nonumber
I_{\alpha }(t) &=& s_\alpha \frac{e\Gamma}{h}  \int d\epsilon_{k} n_{k} (\sum_{\alpha'} \Gamma
|\xi^{(1)}_{k\alpha'}|^2 \\ && -2 \textbf{Im}( \xi^{(1)}_{k\alpha} e^{-2Ni\epsilon_{k}T - i\epsilon^a_{k\alpha} \tau} ) ).
\end{eqnarray}
where  $\xi^{(1)}_{k\alpha}=D^a_{k\alpha} \gamma_N(k) e^{-\Gamma \tau}+ e^{2Ni \epsilon_{k} T}W^a_{k\alpha} (\tau) $. 
$n_k$ is the Fermi-Dirac distribution function. In the sequel we will always
specialize to the zero temperature case ($n_k=1$ for $k<0$, $n_k=0$ for $k\geq 0$). 
In the second half period the current evaluates to
\begin{eqnarray}\label{current_b}\nonumber
I_{\alpha }(t) &=& s_\alpha \frac{e\Gamma}{h}\int d\epsilon_{k} n_{k}( \sum_{\alpha'} \Gamma |\xi^{(2)}_{k\alpha'} |^2 \\ && -2 
\textbf{Im}( \xi^{(2)}_{k\alpha }e^{-2Ni\epsilon_{k}T - i\epsilon^a_{k\alpha} T-i\epsilon^b_{k\alpha} \tau} ) )
\end{eqnarray}
where $\xi^{(2)}_{k\alpha} = D^a_{k\alpha} \gamma_N(k) e^{-\Gamma (T+\tau)}+ e^{2Ni \epsilon_{k} T} (W^a_{k\alpha}e^{-\Gamma \tau}
+ e^{i\epsilon^a_{k\alpha}T} W^b_{k\alpha}(\tau) )$. 
To simplify notation in lengthy expressions we will frequently employ $\Gamma$ as the unit of energy and current, 
and $1/\Gamma$ as the unit of time. In the final results we always reintroduce 
all dimensionful parameters.

\section{Buildup of the steady state}

There is a transient time regime 
after the coupling of the dot to the leads is switched on at time $t=0$
before a steady state has built up.
Initially, the left lead current is opposite to the right one and the initially empty dot is being charged. 
We will see that these transient effects decay exponentially
(proportional to $e^{-\Gamma t}$) to the steady state. 

Let us explicitly look at the two limits of period $T\to \infty$ (dc bias) 
and $T\to 0$ (very fast driving). For $T\to \infty$ one finds from Eq.~(\ref{current_a})
\begin{eqnarray}\nonumber
I_{\alpha }(t) &=& s_\alpha \frac{e}{h} \int d\epsilon_{k} n_{k}
(  \sum_{\alpha'} \frac{1+e^{-2 t} -e^{i\epsilon^a_{k\alpha'}t-t} - e^{-i\epsilon^a_{k\alpha'}t-t} }{
(\epsilon^{a}_{k\alpha'})^2+1} \\ && - 2 
\textbf{Im}[\frac{1-e^{-i\epsilon^a_{k\alpha}t-t}}{\epsilon^a_{k\alpha}-i}]).
\end{eqnarray}
The steady limit ($t\rightarrow\infty$) is 
\begin{equation}
I=\frac{e\Gamma}{h} \int d\epsilon (n(\epsilon+\frac{eV}{2})- n(\epsilon-\frac{eV}{2}))\frac{\Gamma}{\epsilon^2+\Gamma^2}
\end{equation}
which of course coincides with the well-known result for the stationary dc-current \cite{wingreen94},
e.g. for zero temperature
\begin{equation}
I=\frac{2e\Gamma}{h}\,{\rm arctan}\left(\frac{eV}{2\Gamma}\right) \ .
\label{dccurrent}
\end{equation}

In the fast driving limit $T \to 0$ we keep $t=2NT$ invariant and let $N\to \infty$. 
According to the Trotter formula, the evolution then becomes equivalent
to zero voltage bias \cite{Eisler}, 
$\lim_{T\to 0} (e^{iH_a T}e^{iH_b T})^N= e^{i(H_a+H_b)TN}$. 
We find
\begin{eqnarray}\label{highfrequency}
I_{\alpha }(t) = s_\alpha \frac{2e\Gamma e^{-t}}{h} \int d\epsilon n(\epsilon) \frac{e^{-t} - \cos \epsilon t - \epsilon \sin \epsilon t}{\epsilon^2+1}.
\end{eqnarray}
In Fig.~2 we show the transient currents in the left and right lead for different periods $T$ when $\mu=0$. The current oscillations are suppressed when the frequency goes to infinity. 
The $I(t)$-curves gradually change from the dc limit to the high frequency limit described by Eq.~(\ref{highfrequency}) when the period $T$ decreases. 
In the fast driving limit the left and right currents are opposite to each other and both decay to zero with increasing time. 
\begin{figure}
\begin{center}
\includegraphics[width=0.45\textwidth]{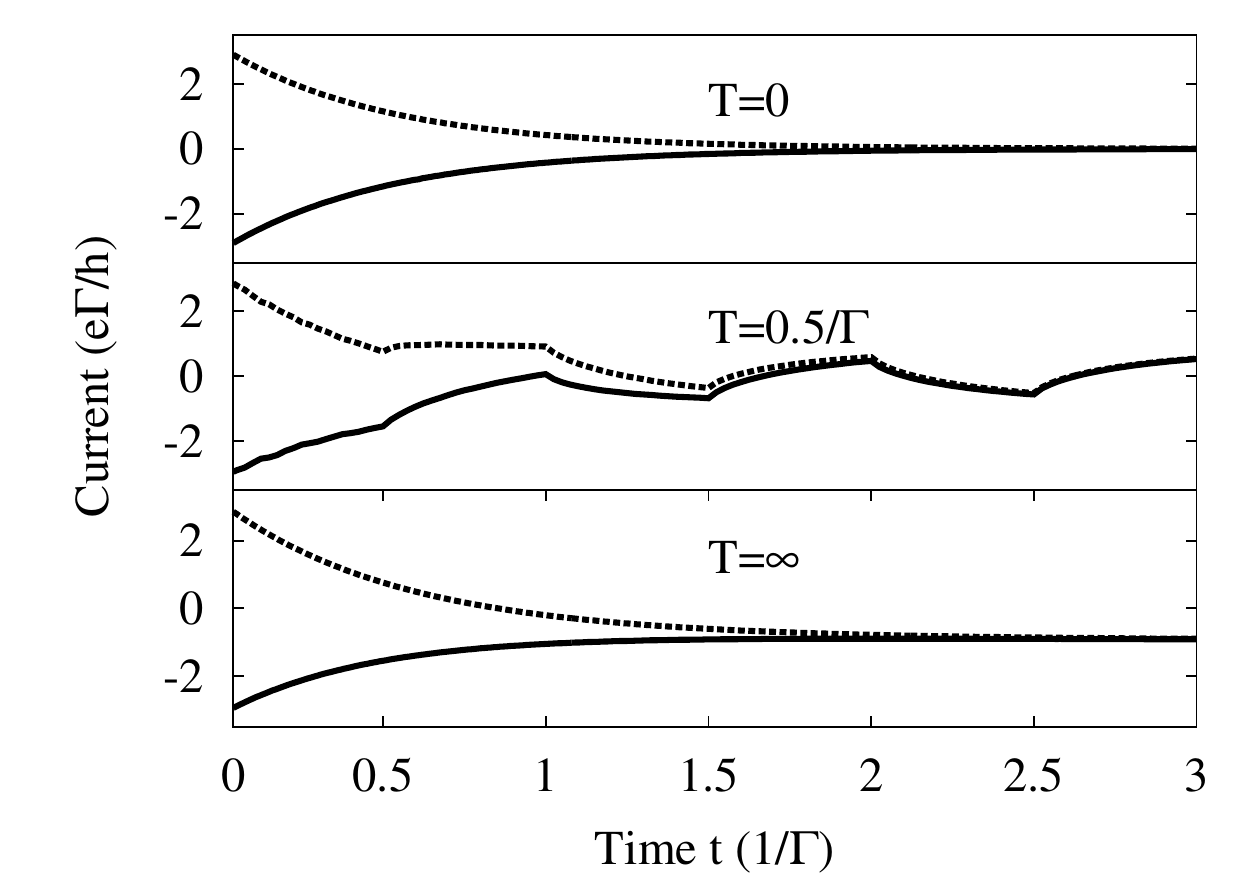}
\caption{Time-dependent current for different switching periods $T$ of the ac voltage bias
(top: infinitely fast driving, middle: intermediate fast driving, bottom: dc case). 
Zero temperature and ac voltage bias $V=\Gamma$ in all graphs. 
The full lines denote the left lead current, the dashed lines the right lead current. 
The hybridization is switched on at time $t=0$.
Notice the discontinuous onset of the current at $t=0$,
which is due to the wide band limit for the conduction band (a detailed discussion can be found in Ref.~\cite{schmidt}).}
\end{center}
\end{figure}

\section{Steady state behavior}

When the time is much larger than $1/\Gamma$, the current reaches its steady state behavior. By taking $N\to \infty$ we 
find this steady state limit given by
\begin{eqnarray}
 I_\alpha(\tau)=  s_\alpha \frac{e\Gamma}{h} \int d\epsilon_k n_k (|\tilde \xi_{k L}|^2+|\tilde \xi_{k R}|^2-2 \textbf{Im} \tilde \xi_{k \alpha} ),
\end{eqnarray}
where $0\leq \tau \leq T$. In the first half period we have 
\begin{eqnarray} \nonumber
\tilde \xi_{k\alpha} = \frac{1}{\epsilon^a_{k\alpha} -i}+\frac{s_\alpha V (e^{2i\epsilon T-i\epsilon^a_{k\alpha}\tau-\tau}-e^{i\epsilon^a_{k\alpha}(T-\tau)
-T-\tau})}{(e^{2i\epsilon T}-e^{-2T})(\epsilon^a_{k\alpha}-i)(\epsilon^b_{k\alpha}-i)}, 
\label{eqfirsthalf} \\
\end{eqnarray}
and in the second half period
\begin{eqnarray}\nonumber
\tilde \xi_{k\alpha} = \frac{1}{\epsilon^b_{k\alpha} -i}+\frac{s_\alpha V (e^{i\epsilon^b_{k\alpha}(T-\tau)
-T-\tau}-e^{2i\epsilon T-i\epsilon^b_{k\alpha}\tau -\tau})}{(e^{2i\epsilon T}-e^{-2T})(\epsilon^a_{k\alpha}-i)(\epsilon^b_{k\alpha}-i)} . 
\label{eqsecondhalf} \\
\end{eqnarray}
From Eqs.~(\ref{eqfirsthalf}) and (\ref{eqsecondhalf}) one immediately verifies that
the steady state current satisfies $ I_\alpha(\tau)= - I_{\bar \alpha}(\tau+ T)$ as expected intuitively, 
where $\bar \alpha$ denotes the opposite lead.

\subsection{Linear response regime}

\begin{figure}
\begin{center}
\includegraphics[width=0.45\textwidth]{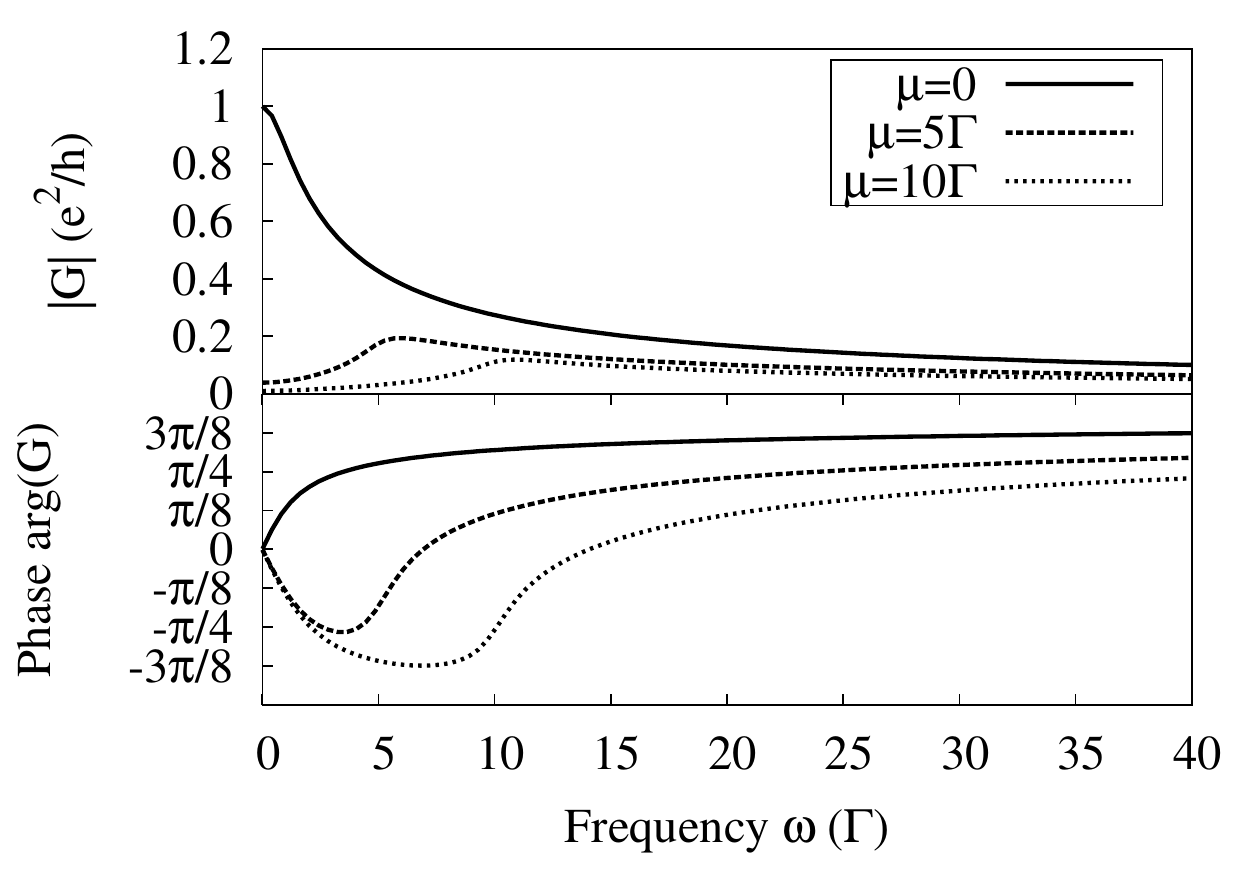}
\caption{The linear admittance of a resonant level model for various level positions
$\mu$ (energy of the dot level with respect to the Fermi energy in the leads) at zero temperature.
The top graph shows the absolute value of the admittance, the bottom one its phase.}
\end{center}
\end{figure}
In the linear response regime of small voltage bias a sinusoidal signal drives a sinusoidal current with the same frequency, and signals with different frequencies can be superimposed linearly. 
Therefore we can factorize the rectangular signal into a series of sinusoidal components and find the 
frequency-dependent complex admittance of the system. 

In the linear response regime the left lead current is equal to the right lead one and can be expressed as
\begin{eqnarray}
\lim_{V\to 0} \frac{I (\tau)}{V}= \frac{e^2}{h} \int d\epsilon n(\epsilon) T(\epsilon),
\end{eqnarray}
where
\begin{eqnarray}
T(\epsilon) =\frac{2\epsilon \Gamma^3}{(\epsilon^2+\Gamma^2)^2} - \textbf{Im} [\frac{2 \Gamma^2 e^{i\epsilon T-i \epsilon \tau -\tau}}
{(e^{i\epsilon T}+e^{-T})(\epsilon-i\Gamma)^2} ].
\end{eqnarray}
We Fourier transform both the ac-voltage signal and the current. We define $I(\omega_n) = \int^{2T}_0 dt e^{i\omega_n t} I(t) = 2\int^T_0 d\tau 
e^{i\omega_{n} \tau} I (\tau)$, where we use the property $I (\tau+T)=-I (\tau)$, and $V(\omega_n) = \int^{2T}_0 dt e^{i\omega_n t} V(t)$. 
Here $\omega_n = \frac{n\pi}{T}$ and $n$ is an odd number. The voltage bias is $-V $ for $0\leq t\leq T$ and $V$ for $T\leq t\leq 2T$, 
leading to $V(\omega_n)= \frac{4V}{i \omega_n}$. By adjusting $T$ the frequency $\omega_n$ can be an arbitrary real number, 
and the linear response admittance $G(\omega)= I(\omega)/V(\omega)$ at zero temperature is given by
\begin{widetext}
\begin{eqnarray}
 G(\omega)= \frac{e^2}{h} \left(\frac{{\rm arccot} \frac{-\omega -\mu}{\Gamma} - {\rm arccot} \frac{\omega-\mu}{\Gamma}}{2 \omega/\Gamma}- 
\frac{i\Gamma}{4\omega} \ln \frac{(\mu^2+\Gamma^2)^2}
{((\mu+\omega)^2+\Gamma^2)((\mu-\omega)^2+\Gamma^2)}\right),
\label{Gomega}
\end{eqnarray}
\end{widetext}
where $\mu$ denotes the position of the dot level with respect to the average Fermi energy of the leads, 
see Fig.~1. Eq.~(\ref{Gomega}) agrees with previous ac-calculations in the linear response regime, see Ref.~\cite{fu}.
Fig.~3 depicts $G(\omega)$ for different level positions $\mu$.
The admittance goes to zero for fast driving, $\omega\to \infty$. For $\omega \to 0$ one recovers 
the well-known dc-conductance $G=\frac{e^2}{h}\frac{\Gamma^2}{\mu^2+\Gamma^2}$. 
For asymmetric dot positions the resonance peak is around $\omega=\mu$, showing the PAT (photon assisted tunneling) effect~\cite{kouwenhoven94}:
When the frequency of the ac-signal is equal to the energy difference of the dot level from the Fermi energy in the leads, 
electrons in the leads can absorb a photon and jump into the dot. Notice from Fig.~3 that the symmetric dot always acts like an inductor as already explained in Ref.~\cite{fu}.
For asymmetric dots there is a crossover from capacitive to inductive behavior around $\omega = \mu$~\cite{fu}. 

\subsection{Beyond the linear response regime}

\begin{figure}
\begin{center}
\includegraphics[width=0.45\textwidth]{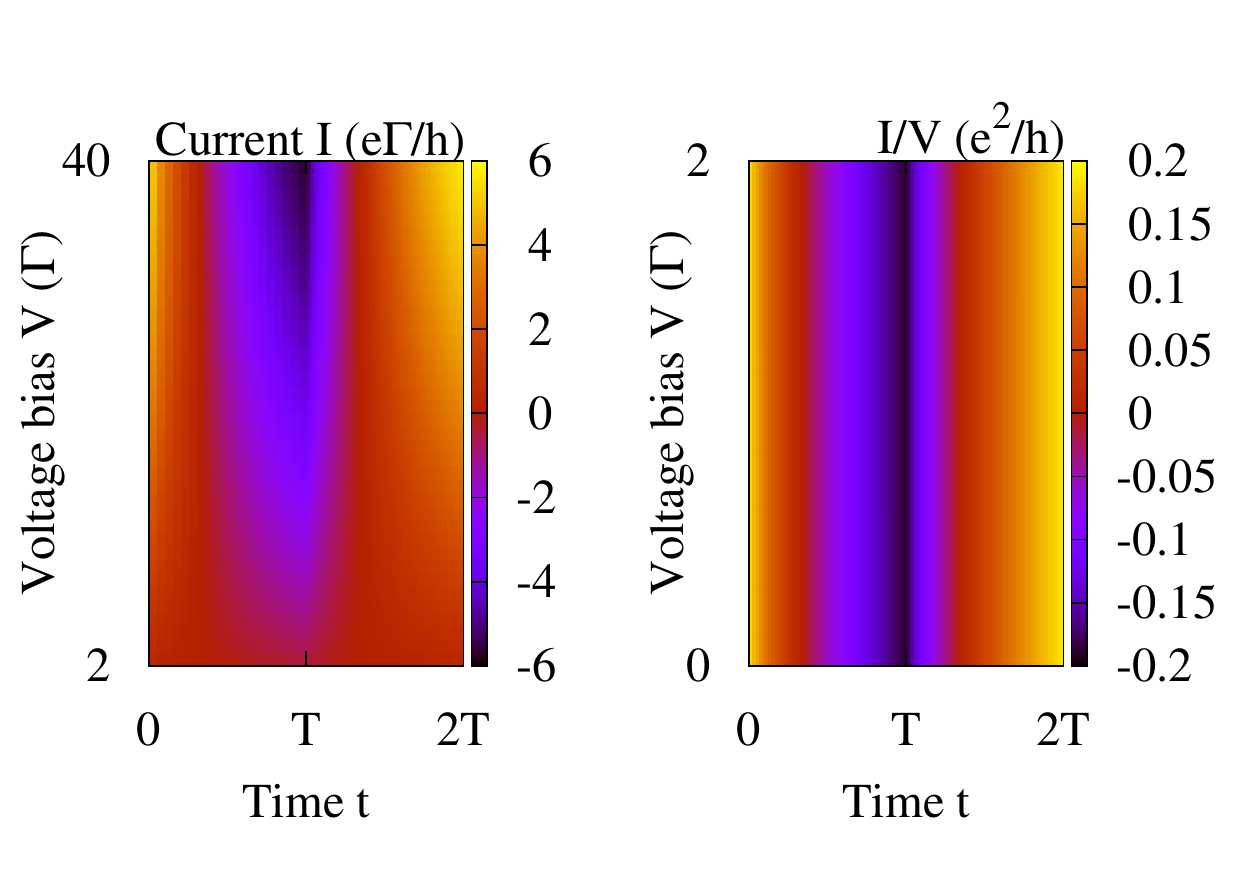}
\caption{The steady state current in one period for fast driving (here $T=0.1/\Gamma$ and zero temperature) in
a symmetric resonant level model ($\mu=0$). The left figure depicts the current~$I$ 
(units $e\Gamma/h$) in 
the nonlinear regime, and the right figure shows $I/V$ (units $e^2/h$) 
for smaller voltage bias  (linear response regime).
Because the driving period is shorter than the time required to establish stationarity in
one period, the time-dependent current looks triangular.}
\end{center}
\end{figure}
\begin{figure}
\begin{center}
\includegraphics[width=0.45\textwidth]{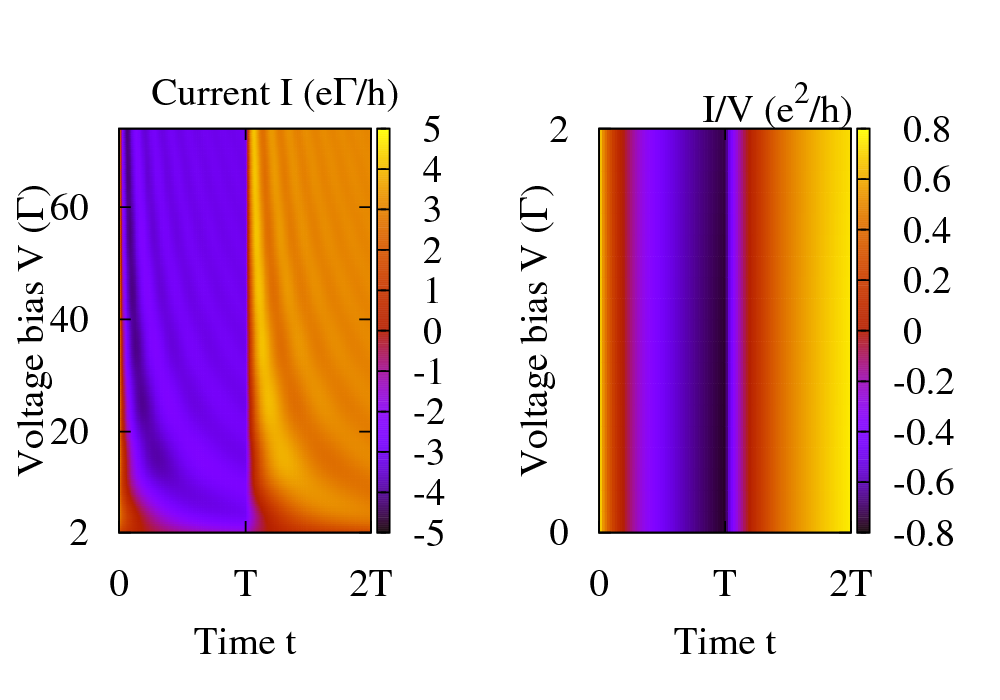}
\caption{The steady state current in one period for intermediate driving (here $T=1/\Gamma$ and zero temperature) in
a symmetric resonant level model ($\mu=0$). The left figure depicts the current~$I$ 
(units $e\Gamma/h$) in 
the nonlinear regime, and the right figure shows $I/V$ (units $e^2/h$) 
for smaller voltage bias (linear response regime).
The oscillations of the current with period $4\pi/V$ ("current ringing" \cite{wingreen94}) 
are clearly visible for large bias.}
\end{center}
\end{figure}
\begin{figure}
\begin{center}
\includegraphics[width=0.45\textwidth]{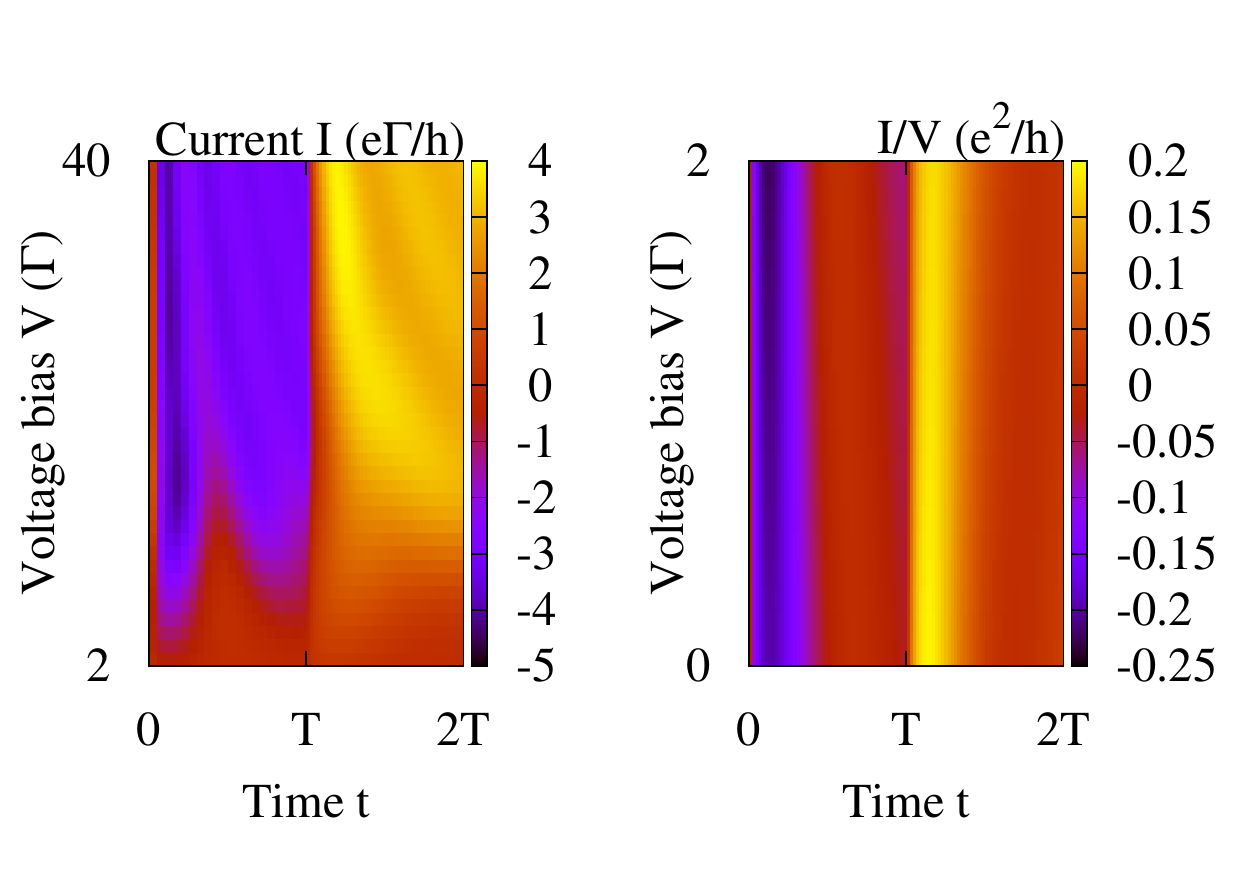}
\caption{
The steady state current in one period for intermediate driving (here $T=1/\Gamma$ and zero temperature) in
an asymmetric resonant level model ($\mu=5\Gamma$). The left figure depicts the current~$I$ 
(units $e\Gamma/h$) in 
the nonlinear regime, and the right figure shows $I/V$ (units $e^2/h$) 
for smaller voltage bias 
(linear response regime). The crossover from capacitive to inductive response
(compare Fig.~3) leads to a complicated behavior of the current in the first half period.}
\end{center}
\end{figure}

For a voltage bias beyond the linear response regime 
it is impossible to calculate $G(\omega)$ by performing a Fourier transformation since the different 
frequency components interact with each other nonlinearly. Therefore we now depict the behavior of the
current $I(t)$ as a function of time~$t$ during one full period in the steady state situation. Due to the
nonlinearities we need to discuss this separately for different driving periods~$T$. We will always
take zero temperature in the sequel, the generalization to nonzero temperature is straightforward.

We first look at fast driving, $T\ll \Gamma^{-1}$.
For the symmetric situation the $I-t$ curve becomes triangled: The current decreases from maximum to minimum in the first half period, and then increases 
from minimum to maximum in the second half period, see Fig.~4. 
In the opposite slow driving limit $T \gg \Gamma^{-1}$, the $I-t$ curve becomes rectangled. The saturated current 
in each half period is simply given by the 
corresponding steady dc-current (\ref{dccurrent}). 
For intermediate driving speed, $T\sim \Gamma^{-1}$, we observe ringing oscillations~\cite{wingreen94} of the current with period $4\pi/V$ (see Fig.~5). 

\begin{figure}
\begin{center}
\includegraphics[width=0.45\textwidth]{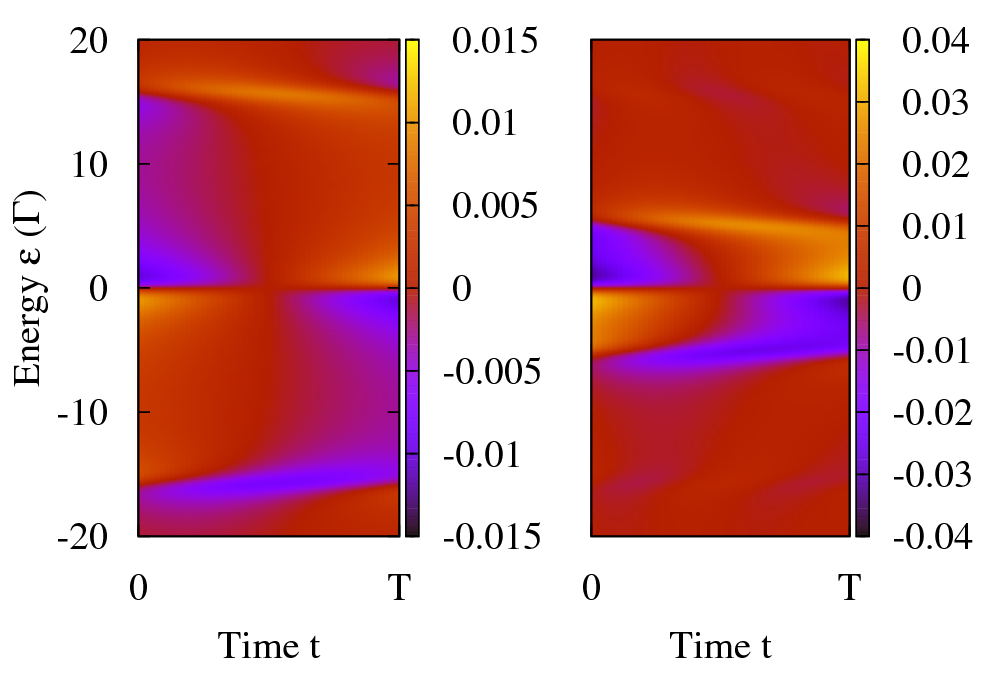}
\caption{The gate differential conductance $G^{\rm gate}_L(\epsilon,\tau)$ (units $e^2/h$) in the linear response regime 
($V=0.2\Gamma$ and zero temperature) for period $T=0.2/\Gamma$ in the left figure and $T=0.6/\Gamma$ in the
right figure.
The pair of bright lines symmetric to $\epsilon=0$ are the PAT lines
at $\epsilon=\pm \pi/T$.}
\end{center}
\end{figure}
For asymmetric dot positions $\mu\neq 0$ the current also has characteristics of PAT and ringing, which
are, however, not easily visible in a plot like Fig.~6. Clear signatures can be found in the 
the differential conductance with respect to the gate voltage, which we denote as 
{\em gate differential conductance}~$G^{\rm gate}$ to distinguish it from the usual definition of
differential conductance with respect to the voltage bias between the leads. We define
\begin{equation}
G^{\rm gate}_\alpha(\epsilon,\tau)\stackrel{\rm def}{=}\frac{d I_\alpha(\tau)}{d \mu}|_{\mu=\epsilon}
\end{equation}
and the current can then be expressed as $I_\alpha (\tau)= \int^\mu_{-\infty} d\epsilon G^{\rm gate}_\alpha(\epsilon,\tau)$. 

\begin{figure}
\begin{center}
\includegraphics[width=0.45\textwidth]{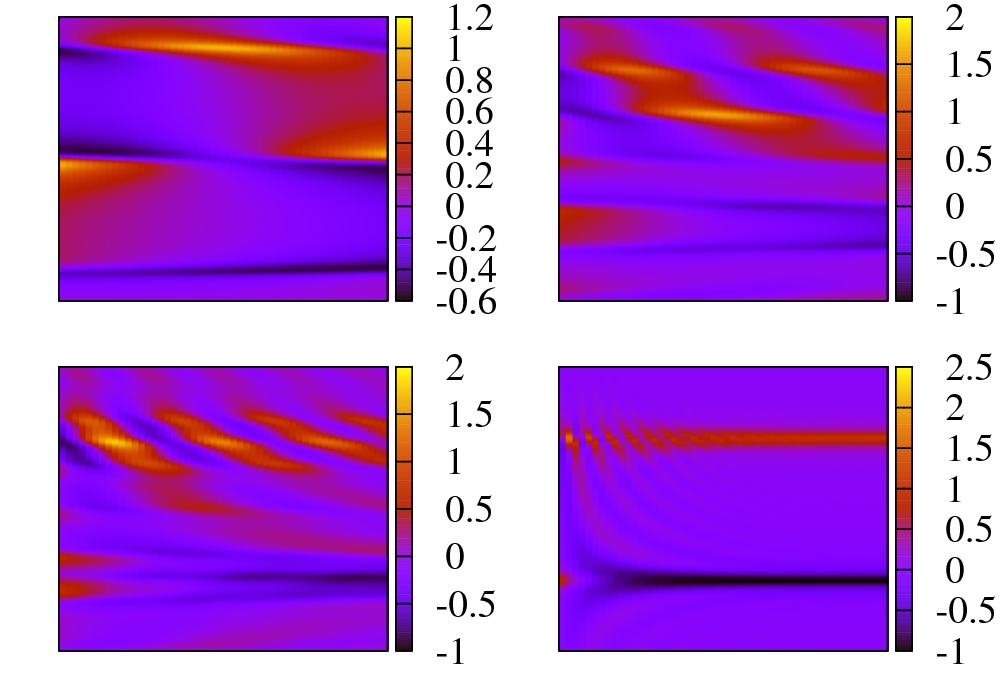}
\caption{The gate differential conductance $G^{\rm gate}_L(\epsilon,\tau)$ 
(units $e^2/h$) for large voltage bias ($V=20\Gamma$) and zero temperature.
The top left figure shows fast driving ($T=0.2/\Gamma$), 
$T=0.5/\Gamma$ in the top right figure, intermediate driving ($T=1/\Gamma$) bottom left
and slow driving ($T=5/\Gamma$) bottom right.
The y-axis denotes the energy ranging from $-20\Gamma$ to $20\Gamma$.
For fast driving ($T\lesssim 0.8/\Gamma$) the higher-order PAT lines are clearly visible.
For slower driving ($T\gtrsim 0.8/\Gamma$) the PAT lines away from $\epsilon=\pm V/2$ disappear with 
increasing~$T$.}
\end{center}
\end{figure}

Figs.~7 and~8 shows $G^{\rm gate}$ in the first half period ($G^{\rm gate}$ in the second half period follows via $G^{\rm gate}_{\rm 2nd}(\epsilon)=G^{\rm gate}_{\rm 1st}(-\epsilon)$).
In the linear response regime we find a pair of bright PAT lines at $\epsilon= \pm \pi/T$ (see Fig.~7). In the regime far from equilibrium,
high order PAT lines at $\epsilon=n\pi/T$ ($|n|\geq 2$) can be observed (see Fig.~8), indicating multiple photon assisted tunneling processes. 
These PAT lines combine and are replaced by a pair of bright resonance lines at $\epsilon=\pm V /2$ when the period increases.
This demonstrates that ac transport for high frequencies is dominated by photon assisted tunneling, and by resonance tunneling for low frequencies.

\section{Conclusions}
We have investigated a resonant level model driven
by rectangular ac-bias in and beyond the linear response regime. 
Even this simple model shows surprisingly rich behavior in its transport properties.
One can observe specific nonequilibrium effects like the buildup of
the steady state, current ringing
and photon assisted tunneling, and the crossover to the well-studied
limiting cases of dc-bias and linear response regime. The results are
exact and based on an explicit diagonalization of the Hamiltonian in
the first and second half period of the rectangular voltage bias driving. 
Within the flow equation framework, this approach can easily be 
generalized to an interacting quantum impurity model exposed to
ac-driving beyond the linear regime. Much less is known about such
systems, which provides another motivation for this work and will
be studied in a subsequent publication.

We acknowledge support through SFB 484 of the Deutsche Forschungsgemeinschaft, the Center for NanoScience (CeNS) Munich, and the German Excellence Initiative via the Nanosystems Initiative Munich (NIM).


\begin{references}
\bibitem{AndersPRL101} F. B. Anders, Phys. Rev. Lett. {\bf 101}, 066804 (2008); J. Phys.: Cond. Matter {\bf 20},  195216 (2008).
\bibitem{Schmidt2008} T. Schmidt, P. Werner, L. M\"uhlbacher, and A. Komnik, Phys. Rev. B {\bf 78}, 235110 (2008).
\bibitem{Millis2009} P. Werner, T. Oka, and A. J. Millis, Phys. Rev. B {\bf 79}, 035320 (2009).
\bibitem{WhitePRL2004} S. R. White and A. E. Feiguin, Phys. Rev. Lett. {\bf 93}, 076401 (2004).
\bibitem{Schmitteckert2004} P. Schmitteckert, Phys. Rev. B {\bf 70},  121302 (2004).
\bibitem{Fuji2003} T. Fujii and K. Ueda, Phys. Rev. B {\bf 68}, 155310  (2003).
\bibitem{Rosch2001} A. Rosch, H. Kroha, and P. Woelfle, Phys. Rev. Lett. {\bf 87}, 156802 (2001).
\bibitem{Schoeller2009} H. Schoeller and F. Reininghaus, Phys. Rev. B {\bf 80}, 045117 (2009); Phys. Rev. B {\bf 80}, 209901(E) (2009).
\bibitem{Kehrein_Kondo} S. Kehrein, Phys. Rev. Lett. {\bf 95}, 056602 (2005); 
P. Fritsch and S. Kehrein, Ann. Phys. {\bf 324}, 1105 (2009).
\bibitem{Meir93} Y. Meir, N. S. Wingreen, and P. Lee, Phys. Rev. Lett. {\bf 70}, 2601 (1993).
\bibitem{nordlander} P. Nordlander, M. Pustilnik, Y. Meir, N. S. Wingreen, D. C. Langreth, Phys. Rev. Lett. {\bf 83}, 808 (1999).
\bibitem{FabianNewJPhys} J. Eckel, F. Heidrich-Meisner, S.G. Jakobs, M. Thorwart, M. Pletyukhov, and R. Egger, Preprint arXiv:1001.3773
\bibitem{tien} P. K. Tien and J. P. Gordon, Phys. Rev. {\bf 129}, 647 (1963).
\bibitem{kouwenhoven94} L. P. Kouwenhoven, S. Jauhar, J. Orenstein, P. L. McEuen, Phys. Rev. Lett. {\bf 73}, 3443 (1994).
\bibitem{wingreen94} A. P. Jauho, N. S. Wingreen, Y. Meir, Phys. Rev. B {\bf 50}, 5528 (1994).
\bibitem{wingreen93} N. S. Wingreen, A. P. Jauho, Y. Meir, Phys. Rev. B {\bf 48}, 8487 (1993).
\bibitem{maciejko06} J. Maciejko, J. Wang, H. Guo, Phys. Rev. B {\bf 74}, 085324 (2006).
\bibitem{Goker08} A. Goker, Solid State Commun. {\bf 148}, 230 (2008).
\bibitem{plihal} M. Plihal, D. C. Langreth, Phys. Rev. B {\bf 61}, R13341 (2000).
\bibitem{nordlander2000} P. Nordlander, N. S. Wingreen, Y. Meir, and D. C. Langreth, Phys. Rev. B {\bf 61}, 2146 (2000).
\bibitem{hershfield96} A. Schiller, S. Hershfield, Phys. Rev. Lett. {\bf 77}, 1821 (1996).
\bibitem{hershfield95} A. Schiller, S. Hershfield, Phys. Rev. B {\bf 51}, 12896 (1995).
\bibitem{hershfield00} A. Schiller, S. Hershfield, Phys. Rev. B {\bf 62}, R16271 (2000).
\bibitem{levelV} For mathematical simplicity we assume that $V/2$ is an integer multiple of the level spacing $\eta$. In other words $\epsilon_k\pm \frac{V}{2}$
can be written as some $\epsilon_{k'}$. This condition will of course play no role in the thermodynamic limit.
\bibitem{Eisler} V. Eisler and I. Peschel, Ann. Phys. (Berlin) {\bf 17}, 410 (2008).
\bibitem{schmidt} T. L. Schmidt, P. Werner, L. M\"{u}hlbacher, A. Komnik, Phys. Rev. B {\bf 78}, 235110 (2008). 
\bibitem{fu} Y. Fu, S. C. Dudley, Phys. Rev. Lett. {\bf 70}, 65 (1993).

\end{references}
\end{document}